\documentclass[conference]{IEEEtran}
\IEEEoverridecommandlockouts
\usepackage{cite}
\usepackage{amsmath,amssymb,amsfonts}
\usepackage{algorithmic}
\usepackage{graphicx}
\usepackage{textcomp}
\usepackage{xcolor}
\def\BibTeX{{\rm B\kern-.05em{\sc i\kern-.025em b}\kern-.08em
    T\kern-.1667em\lower.7ex\hbox{E}\kern-.125emX}}

\usepackage{hyperref} 
\hypersetup{
    colorlinks,
    linkcolor={red!80!black},
    citecolor={blue!80!black},
    urlcolor={blue!80!black}
} 
\usepackage{orcidlink} 
    
\begin{document}

\title{HMAS: enabling seamless collaboration between drones, quadruped robots, and human operators with efficient spatial awareness}

\author{\IEEEauthorblockN{Amaury Saint-Jore \orcidlink{0009-0009-5292-8234}}
\IEEEauthorblockA{\textit{LORIA, CNRS, Université de Lorraine} \\
Nancy, France \\
amaury.saint-jore@loria.fr}
\and
\IEEEauthorblockN{Ye-Qiong Song \orcidlink{0000-0002-3949-340X}}
\IEEEauthorblockA{\textit{LORIA, CNRS, Université de Lorraine} \\
Nancy, France \\
ye-qiong.song@loria.fr}
\and
\IEEEauthorblockN{Laurent Ciarletta \orcidlink{0000-0003-4160-0072}}
\IEEEauthorblockA{\textit{LORIA, CNRS, Université de Lorraine} \\
Nancy, France \\
laurent.ciarletta@loria.fr}
}

\maketitle

\begin{abstract}
Heterogeneous robots equipped with multi-modal sensors (e.g., UAV, wheeled and legged terrestrial robots) provide rich and complementary functions that may help human operators to accomplish complex tasks in unknown environments. 
However, seamlessly integrating heterogeneous agents and making them interact and collaborate still arise challenging issues.
In this paper, we define a ROS 2 based software architecture that allows to build incarnated heterogeneous multi-agent systems (HMAS) in a generic way. We showcase its effectiveness through a scenario integrating aerial drones, quadruped robots, and human operators (see \url{https://youtu.be/iOtCCticGuk}). In addition, agent spatial awareness in unknown outdoor environments is a critical step for realizing autonomous individual movements, interactions, and collaborations. Through intensive experimental measurements, RTK-GPS is shown to be a suitable solution for achieving the required locating accuracy.
\end{abstract}


\section{Introduction}

Nowadays, we observe that more and more robots are operated or considered in different applications.
Tasks, environments, and situations are diversifying and becoming more complex. There is a need of increasingly complex, flexible, and adaptive robotic systems. The use of robots is essential for \textbf{exploration}, \textbf{surveillance}, and \textbf{recognition} as in search and rescue (S\&R) \cite{CMR_SaR}. The primary objective is to help humans in different missions that may involve high risk and require a large operating area. There is a wide variety of robots, drones, rovers, underwater robots, etc. Each with its own advantages on specific topics, but none covers all the application needs.
Solutions and applications are proposed to use a large number of heterogeneous robots at the same time: these are multi-agent systems. Each robot (called an agent) or the system as a whole can be controlled by one or more human operators.
The objective is to have this group collaborate and interact to accomplish together the targeted tasks. In addition, the system is intended to be autonomous in most of its actions, requiring the least human intervention. Although research works on multi-robot systems are progressing significantly, they are not yet widely used in real-world applications, for example in the field of S\&R \cite{drew_multi-agent_2021}. One of the main problems is the high development cost and complexity of integrating heterogeneous software systems with their own data model, locating systems, communication, and programming interfaces.  

\begin{figure}[htbp]
\centering
\includegraphics[scale=0.08]{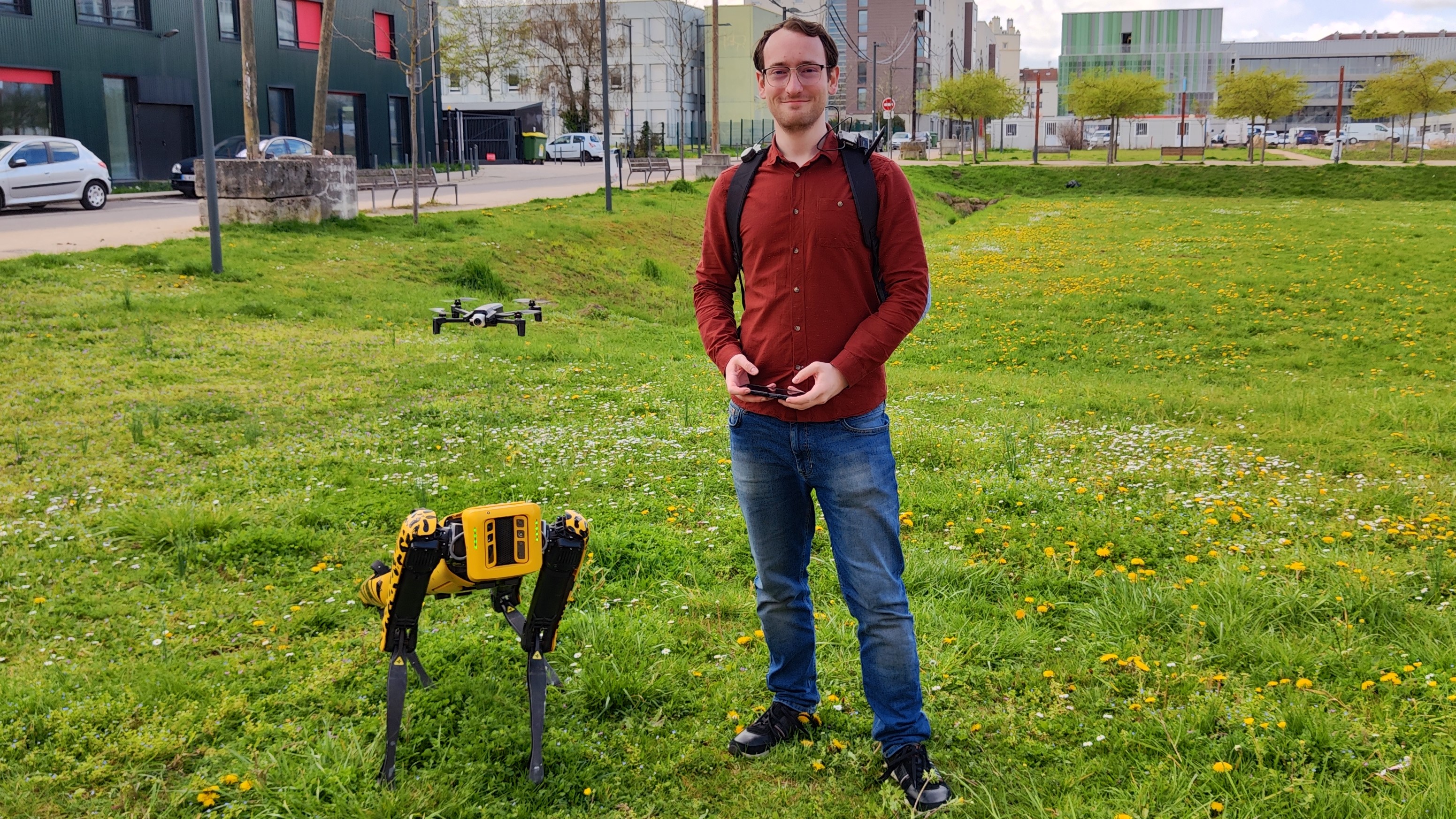}
\caption{An HMAS composed of a ground quadruped robot (Spot), an aerial drone currently flying (Anafi), and a human operator equipped with sensors on a backpack. All agents and humans have RTK GPS to locate themselves in an outdoor environment.} \label{figure_HMAS_syst}
\end{figure}

We propose an operational solution to build a \textbf{multi-agent collaborative system} composed of \textbf{heterogeneous agents}: terrestrial robots, flying drones, and human operators. All the agents of the system can \textbf{collaborate} and \textbf{interact} with each other, but also with physical elements and humans outside the system. 
An implementation of our system has been realized showing robot agents (drone or quadruped) following each other, and also following detected tags or humans located with Real Time Kinematic (RTK) GPS (see \url{https://youtu.be/iOtCCticGuk}).
In this paper, among numerous technical issues, we mainly focus on the integration of heterogeneous robots, and how to correctly carry out both internal and external interactions and collaborations, therefore how to accurately locate them in the environment relatively. Two key issues will be detailed to move towards the autonomous collaboration solution: the creation of an \textbf{HMAS} and the \textbf{geolocation} of agents. 
Many other aspects (e.g. image recognition, wireless communication, robustness, security) are parts of our ongoing work that are not presented in this paper due to the space limitation.
For instance, our HMAS has been used to implement and study a role-based trust model assessing Internet of Agents services \cite{saint-jore}.
Our contribution is three-fold: 
1) define a generic ROS 2 based software architecture allowing to seamlessly build HMAS;
2) validating through experimentation an RTK-GPS based accurate agent locating solution, allowing to realize autonomous and collaborative movement in an unknown dynamic environment;
3) showcased the effectiveness of our proposal through the implementation of an HMAS scenario including aerial drones (Anafi, \texttrademark Parrot) and ground robots (Spot, \texttrademark Boston Dynamics) with operators equipped of sensors Fig. \ref{figure_HMAS_syst}. 
In contrast with many simulation-based existing works, our work goes beyond the simulation step, by implementing and experimenting an HMAS in the \textbf{real world}, both in outdoor and indoor environments. Our agents can evolve in real-world conditions, respecting environmental and safety constraints, in \textbf{unknown environments} including dynamic obstacles and humans.
In this paper, we only deal with outdoor spaces.
For indoor location without GPS, a lot of work has been done \cite{harik_fuzzy_2017,hughes_colony_2014}. Often, this consists of using, for indoor example with HMAS, an inertial unit, a depth camera, or a LiDAR.
In our agents we used LiDAR, from the initial conditions, to determine the robot state frame of each robot in their own reference frame and perform SLAM.

The rest of the paper is organized as follows. Section 2 presents related works. Section 3 describes the design of our ROS 2 based HMAS software architecture. Agent locating issue is addressed in Section 4, together with experimental measurement analysis. Section 5 concludes the paper and points out some future directions.

\section{Related works}

One of the important challenges of an HMAS is the \textbf{heterogeneity}. It is interesting to combine the heterogeneous data received and to take advantage of each type of robot to best accomplish the overall mission. This allows us to have the various advantages of each specific agent and to overcome their limits \cite{hughes_colony_2014}.
However, a number of challenges remain.
In general, and particularly in the S\&R domain, we don't yet know how to easily design and deploy such complex systems \cite{CMR_SaR}. Also, many MAS have been developed only in simulation, with sometimes a controlled real-life environment, but few systems are applied in real-life conditions, and still less combining aerial and ground robots \cite{ch_mrs_survey}, \cite{verma_multi-robot_2021}. Given the complexity of HMAS, these systems often deal with only a part of the problem.
Let's take three examples covering various interesting aspects of HMAS.

In \cite{harik_towards_2016,harik_fuzzy_2017}, a collaboration between a land mobile robot and a drone for the realization of an automated inventory in a logistics warehouse was developed. The idea is to combine two very different modalities of locomotion. Another similar case is a robot dog and drone collaboration which corresponds to our system Fig. \ref{figure_HMAS_syst}, equipped with a rotating LiDAR sensor, and RTK GPS coupled to a drone delivering an aerial view, in a real environment for inspection, payload delivery, S\&R \cite{bellicoso_advances_2018}. They autonomously and semi-autonomously navigate in real-world scenarios to complete high-level tasks. However, it's not easy to achieve individual tasks like mobility, navigation, and inspection, which could be entirely fulfilled, and even less for a heterogeneous system composed of several agents in harsh conditions. It is difficult to obtain a reliable and robust system in a real environment. 

An applied system in the DARPA Subterranean Challenge described in \cite{agha_nebula_2021,bouman_autonomous_2020}, with the goal to develop fully autonomous systems to explore subsurface voids, illustrates well the strong points and the difficulties of the creation of such heterogeneous robotic systems. Accomplishing the objectives of complex real-world operations with constraints on time, resources/cost, robot size, weight, power, etc., can be too difficult or impossible for a single robot, hence the use of a multi-agent system. A DARPA challenge team called CoSTAR has proposed a framework NeBula (Networked Belief-aware Perceptual Autonomy) in \cite{agha_nebula_2021}, applied on a quadruped robot in \cite{bouman_autonomous_2020}, but the associated architecture is neither modular nor adaptive and reproducible. The system building required extensive efforts, involving a lot of staff, payloads on robots, and layers of programming. It is only deployable in closed environments.

Finally, a heterogeneous unmanned swarm system was deployed on a large-scale outdoor environment, composed of ground and aerial robots to map an area and facilitate missions to be carried out \cite{clark_ccast_2021}. There are five important issues addressed: the number of agents, agent complexity, heterogeneity, collective complexity, and human-swarm interaction. Some additional issues were raised during tests. Quickly detecting and positioning all agents to perform tasks is not enough. The operator controlling the swarm system receives too much information because of the large number of agents, and his interaction environment is not designed. The robots in the system lack the sensors, embedded resources, and autonomy necessary for long-term unsupervised autonomy, so it requires constant human attention. Key open issues remain about how to design and build swarm systems suitable for a large range of possible operating environments, in large open spaces where agents can operate at high speed and safely.

Through all these systems, one of the main issues is the creation of the software architecture for seamlessly integrating heterogeneous agents. Several Robotics Software Frameworks  (RSF) exist and provide the basic infrastructure necessary for the development of Multi-Agent Robotic Systems \cite{inigo-blasco_robotics_2012}. Each software has certain characteristics and advantages with respect to the others in several aspects: concurrent and distributed architecture, modularity, robustness and fault tolerance, real-time, and efficiency. One of the best-known and widespread software is Robot Operating System (ROS), used in many research projects. ROS is for example used in a heterogeneous Multi-Agent System composed of mobile robots and a quadcopter \cite{hughes_colony_2014}. The use of a RSF such as ROS depends on the needs and structure of the system. It may be useful for defining the composition of a team in terms of roles describing the types of work normally expected of its members. 
In addition, system dynamics can be implemented in the case of structural changes, loss or failure of an agent, or the discovery of potential new agents. \cite{gunn_dynamic_2015}. This dynamic heterogeneous team formation for robotic urban S\&R shows communication distance and reliability are important. Also, a few ROS 1 based frameworks like CoSTAR \cite{agha_nebula_2021,bouman_autonomous_2020} exist but they are mainly dedicated to a single robotic system and are not well adapted to multi-agent systems, where there is a need for modularity and easy data exchange, for example. In this case, the use of ROS 2 \cite{eros_ros2_2019} could be a solution to this issue.

\begin{figure*}[htbp]
\centering
\includegraphics[scale=0.5]{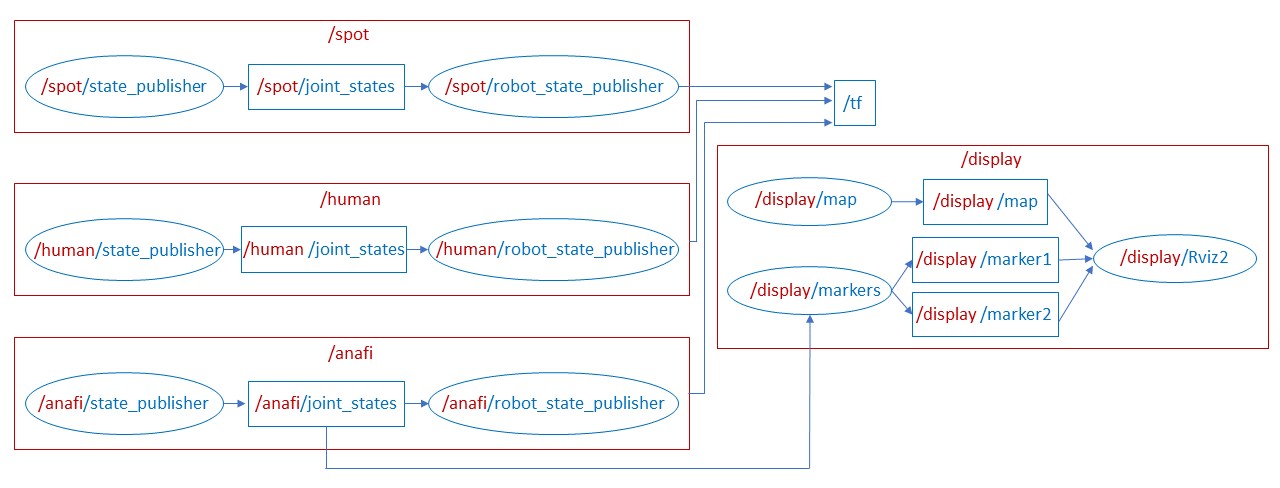}
\caption{Simplified example illustrating the HMAS architecture with the use of namespaces (red) to isolate agents with their nodes (blue circle) and topics (blue square). The outputs of the nodes are publishers, and the inputs are subscribers, so a node can play both roles.}
\label{figure_graph_namespace}
\end{figure*}

\section{Design of a Heterogeneous Multi-Agent System}

After reviewing different systems, we propose the realization of a \textbf{heterogeneous multi-agent system}. The agents will be collaborative with each other to accomplish different tasks, but also in interaction with entities internal and external to the system. The design and construction of an HMAS will be examined first. The development of our solution and the incorporation of various robots and operators as agents of the system will then be covered.

\subsection{HMAS design requirements}

How to design an HMAS in terms of its structure is our first identified issue.
First of all, an HMAS is a system whose first and foremost characteristic is its composition. The system is defined by the agents that are parts of it. Our set of agents is composed of elements of different natures, it is \textbf{heterogeneous}. The system can have several agents, whether similar or not, belonging or not to the same category. The system is also made up of several categories on the ground or in the air, including human operators equipped with the appropriate devices (GPS, communication system, etc.).
Reliability, autonomy, and mobility are three criteria that can be used to generalize a multi-robot system's capability. Perception and planning, communication, human-robot interaction, and cost/function trade-off are the technical difficulties associated to the usage of multi-robot systems for various applications as S\&R \cite{drew_multi-agent_2021}.

To design and develop an HMAS, a wide variety of agents will have to be programmed, using several programming languages, API and SDK. Tasks and calculations have to be executed in parallel and multithreading and multiprocessing have to be performed. It is also necessary to manage the data storage and transfer internally and externally using various means (queue, pipe, socket, etc.). The architecture at hardware and especially software level is therefore complex for developing such an HMAS, but also for the users and operators of the system. This requires the system to be well organized. For multi-agent systems, various tools and RSFs exist including middleware. But one RSF has all the necessary characteristics to realize an HMAS, it is the ROS middleware \cite{inigo-blasco_robotics_2012}.

\subsection{ROS 2 based HMAS}

ROS is an open-source software platform for robotics applications providing tools, libraries, and capabilities to build robot systems \cite{macenski_robot_2022}. To solve many problems linked to the creation of a reliable multi-robot system, a second version ROS 2 was designed, based on a Data Distribution Service (DDS). Thanks to the publisher/subscriber of DDS used for message and data exchange (called topics), there is no need for a master to manage the system, everyone is found on the communication network using a discovery mechanism \cite{eros_ros2_2019}. Components of agents (e.g., functions, sensors, and actuators) are ROS nodes. Even if ROS 2 allows the realization of fleets of drones, there is no standard approach to realize a multi-agent system.

An HMAS will consist of robots either built by oneself or purchased. The advantage of ROS 2 is that packages are already developed to use some commercial sensors and robots allowing quick use. Otherwise, the hardware has to be integrated into ROS 2 and the necessary packages developed, so this middleware is double-edged. Moreover, many packages have not yet migrated to ROS 2, not to mention the compatibility issues between the different software distributions. It is preferable to focus on ROS 2 to avoid the use of a bridge between ROS 1 and 2 and the problems of backward compatibility.
With this middleware, we are not going to encapsulate each function and method of an SDK of a robot or sensors for example, but to obtain nodes where each one carries out a functionality or an action of the system. We can thus execute a node or a set of nodes on several computers that communicate with each other \cite{eros_ros2_2019}. With ROS 2, the software architecture of a system is independent of its hardware architecture. Packages allow the development and reuse of nodes and functionalities, in particular by type of robot, sensors, analysis tools, etc. The nodes are intended to be as generic as possible and reusable using dynamic parameters at the nodes' inputs.

We are now interested in defining agents within ROS 2. The problem here is how to conceptualize an agent in our system architecture. An agent will be composed of many nodes, parameters, topics, services, and actions. However, our system is intended to be multi-agent, so we will find the same composition several times. And it is also heterogeneous, leading to different compositions between types of agents and between agents of the same type according to their payloads and functionalities.
As mentioned above, to build a multi-robot system, there is no standard approach. One method proposed in the state of the art is to improve the communication architecture using socket and built-in features of ROS 2 \cite{dahl_expanding_2021}. The agents are thus separated and placed in different domains, communicating via socket bridges. However, this adds a level of complexity to the system and reduces its performance.
We propose a simpler and more efficient solution using the notion of \textbf{namespace} in Fig. \ref{figure_graph_namespace}. Every object in ROS 2, be it nodes, topics, services and actions has a name. We choose to define these names in a relative way. Thus, when they are used in a launch file, we add a namespace to the nodes, where each namespace corresponds to an agent in the system and therefore to the name of this agent. Two similar robots can therefore use the same architecture but are separated into two spaces. The second point is that some HMAS nodes may need the name of an agent or the namespace in which they are executed. We, therefore, pass the name of the agent that corresponds to the desired namespace as an input parameter to the node when it is initialized. Our HMAS thus has a modular and adaptive architecture for the use of several heterogeneous agents.

\begin{figure}[htbp]
\centering
\includegraphics[scale=0.65]{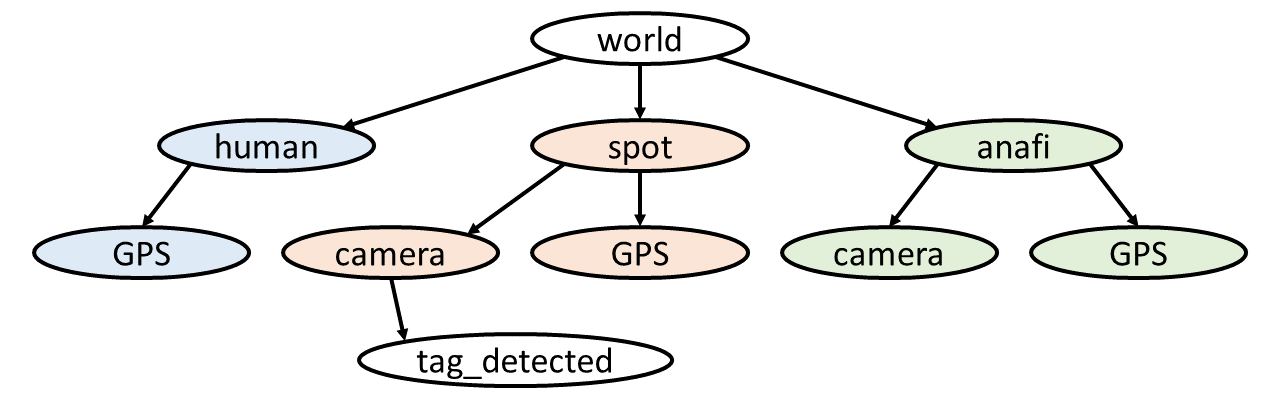}
\caption{Simplified transformation tree with three agents equipped with various sensors (e.g., camera, GPS). In this way, the relationships between frames can be easily defined.} \label{frames}
\end{figure}

\subsection{Development and integration}

We will now develop our system and integrate different ground and air robots to create an example of HMAS. For our first tests, we limit ourselves to the use of three agents, a quadruped robot Spot built by Boston Dynamics, an Anafi drone from Parrot, and an operator equipped with sensors as in Fig. \ref{figure_HMAS_syst}. Our various agents are equipped with RTK GPS, which we will discuss in the next section, to locate their positions. Agents in our system communicate centrally via wifi: all devices are connected to a single computer. But our generic architecture design allows us to easily replace Wifi by other network protocols such as 4G or 5G if a larger area needs to be covered.
Of course, to perform the initial move tests and to check the functioning of the system, especially for aerial drones, we use the APIs provided with Gazebo and Unreal Engine to first simulate and then use the real drone. Whether it's simulation or real testing, it doesn't change the way ROS 2 and our HMAS work and are used.

ROS 2 offers us a wide range of available and reusable tools to develop and operate our system. Thanks to the graphical user interface of ROS 2 called RQt, we can visualize our system in a node graph composed of nodes, topics, and namespaces as shown in Fig. ~\ref{figure_graph_namespace}. In addition, in our HMAS there are many different coordinate systems called frames, representing for example robot states or sensors. Thanks to RQt, we obtain the transformation tree (TF tree) describing the frame transformations between the different agents of the system and their world reference (Fig. \ref{frames}). Finally, we can dynamically configure the system parameters, and plot the graphs to visualize the evolution of these parameters with the plot tool. Using  ros2bag, we record published data on topics in our system and can replay the data to reproduce the results of tests and experiments.

\begin{figure*}[htbp]
\centering
\includegraphics[scale=0.4]{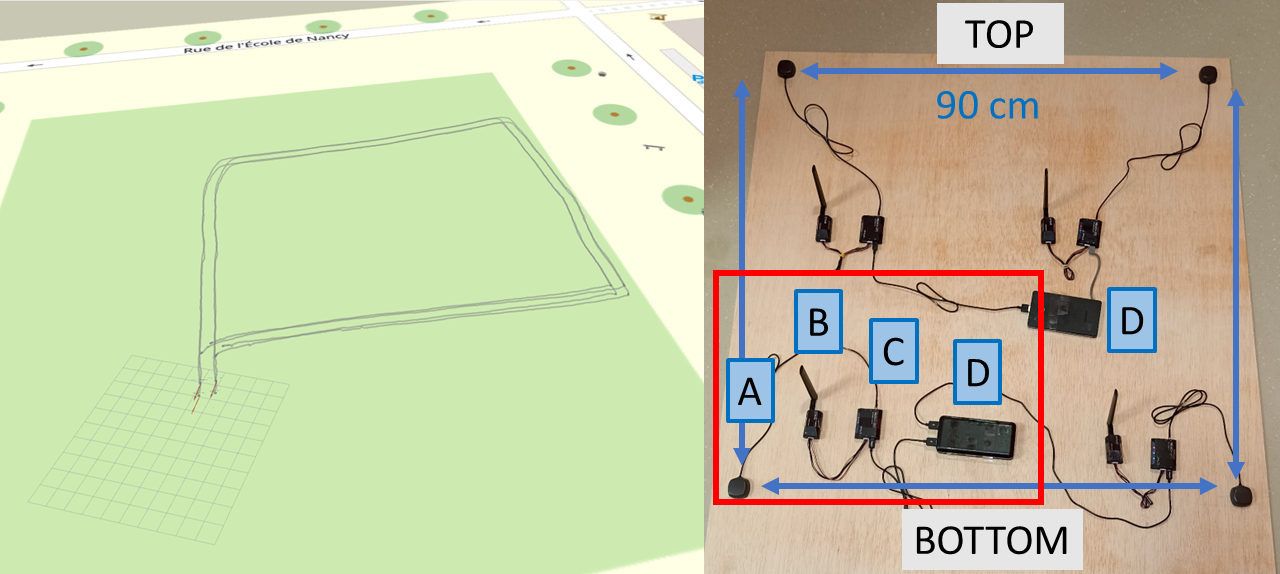}
\caption{\textbf{(a)} RViz 2 to represent the 3D displacements of the agents on a map. We can see the trails of the motions but also the agents with 3D models. The base (the origin) is located in the center of the grid. One square of the grid represents 1 meter.
\textbf{(b)} A rover (red square) is composed of its GPS antenna (A), its LoRa antenna (B) to receive the correction, and the RTK corrector (C). They are powered by two batteries (D). The base (not visible here) is made of the same hardware and placed near the board.} \label{rviz_carre}
\end{figure*}

To visualize our system in 3D, we use the RViz 2 tool (Fig. \ref{rviz_carre}.a), generated by the observer (display) in the node graph. The different agents of our system including operator and ground, air and robots can be modeled, as well as the external and internal environment with the help of a world map and a plan.
Concerning the quality of service of the data exchanged within ROS 2, we consider that the topics require real-time features, we use a best effort reliability policy, volatile durability policy, and keep last history policy with the queue size of one \cite{ginting_chord_2021} so that the freshest data is always used.

Different components have been implemented making the system fully operational (\url{https://youtu.be/iOtCCticGuk}).
Thanks to the modularity and flexibility of this ROS 2-based system, we can easily use it to study various aspects of HMAS. For example, our system has already enabled us to carry out a study of trust in our HMAS \cite{saint-jore}.
In this paper, we only detail one aspect of a multi-agent system, the geolocation which is an important key step for any outdoor HMAS.

\section{Locating Agents in HMAS}

After having defined the ROS 2 based software architecture, we wish to carry out interactions and collaborations between the robot and human agents, both internal and external to the system. For example, a robot can follow a human on a path by staying next to him. This requires the spatial awareness of agents. To do this, a critical issue is to localize the system's agents as precisely as possible \textbf{with centimeter accuracy}, to be on a human scale i.e. below 25 cm. We will therefore see how to use RTK GPS to locate our agents in the environment and realize experiments to test the accuracy and efficiency of such a device.

\subsection{Geolocation with RTK GPS}

In order to carry out the movements of our agents and in particular interactions and collaborations, we need to accurately locate them in space.

This is more complicated when using several different entities, especially if they are not initialized at the same time to operate in the same environment.
In an uncontrolled dynamic outdoor space, we can use a geolocation system. Our goal is that the different robot and human agents are expressed in the same transformation tree, with the same world frame (see Fig. \ref{frames}). To position our agents within a few meters we can use conventional GPS, but we need higher accuracy to perform motions and collaborations. In addition to the classical method of using IMU, LiDAR, or other data to perform data fusion, we propose to use \textbf{Real Time Kinematic (RTK)} with Global Navigation Satellite Systems (GNSS).


\begin{figure*}[htbp]
\centering
\includegraphics[scale=0.5]{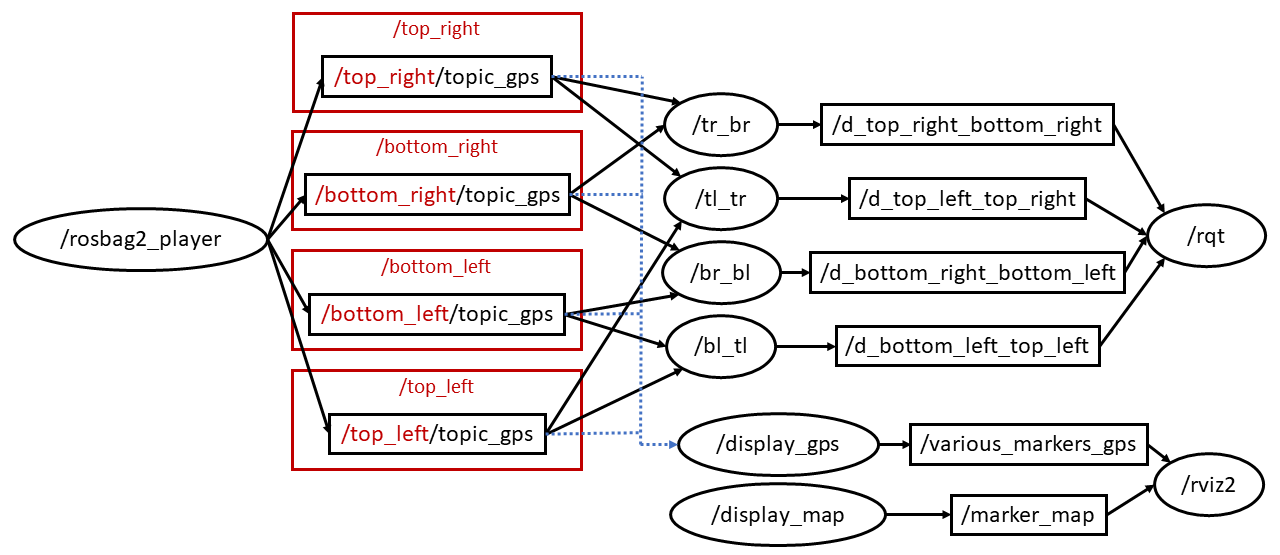}
\caption{ROS 2 system used in the experiments to retrieve data, plot it and display it in 3D. We have the four GPS fixed in the four sides of the board.} \label{rosgraph_carre}
\end{figure*}

The principle is to use a fixed \textbf{GPS base} to send the correction to different mobile GPS called \textbf{rovers}
\cite{noauthor_docs_nodate,bellicoso_advances_2018}. RTK is mainly used in autonomous vehicles and agriculture with expensive hardware and subscription. We use \texttrademark Emlid Reach RTK GPS for the base and rovers which are cheap and easily integrated on our agents (Fig. \ref{rviz_carre}.b).
The coverage of the base to send the correction is 10 to 60 km in LoRaWAN with the hardware used. To further extending coverage one can use a NTRIP caster to pass RTK corrections between the receivers through the internet \cite{noauthor_docs_nodate}. It is also possible to use one's own base, which we will do. Note that it is possible to use a private network of bases from a company, which is very expensive, or else to use a network accessible for free, for example in France \cite{julien_jancelindocs-centipedertk_2022}, which is deployed throughout the whole territory. Also, a rover is not limited to a single base, one can jump from one to another. Thus RTK technology is applicable everywhere and is almost not limited by a range.
In addition to locating our agents in our system with high accuracy, the second advantage of RTK GPS is that it allows us to define a common world frame for all our agents in our HMAS under ROS 2. We use our RTK base as the reference frame for our world frame. We can thus use the East North Up (ENU) Cartesian coordinate system. This eliminates the need to represent our environment using a geodesic system (latitude, longitude, altitude) which is complicated to manipulate for 3D movements. Thus our agents do not evolve in a simulation or in a controlled environment, but in a representation of the real world in an unknown (unexplored) and dynamic environment. We can therefore represent our agents in 3D in RViz as well as the map of the real world from the OpenStreetMap data (Fig. \ref{rviz_carre}.a).

So our solution for locating the agents in our system involves using RTK GPS for precise positioning, being able to express them in a common reference frame defined by a fixed GPS base, and finally representing them in 3D in the real environment with an ENU Cartesian reference frame to simplify their movements.

\subsection{Experimentation with RTK GPS accuracy}

\begin{figure}[htbp]
\centering
\includegraphics[scale=0.4]{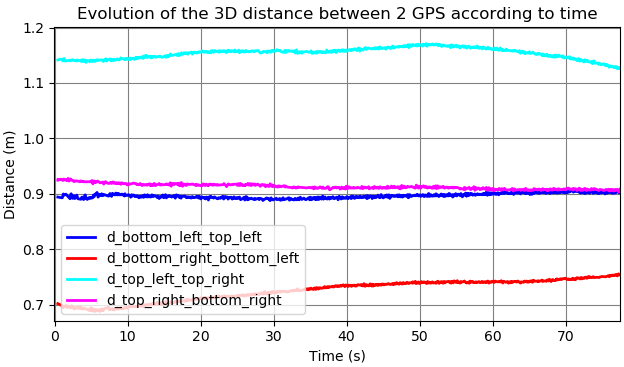}
\caption{Distances between 4 RTK-GPS: case without motion and without disruptions.} \label{plot sur place}
\end{figure}

\begin{figure}[htbp]
\centering
\includegraphics[scale=0.4]{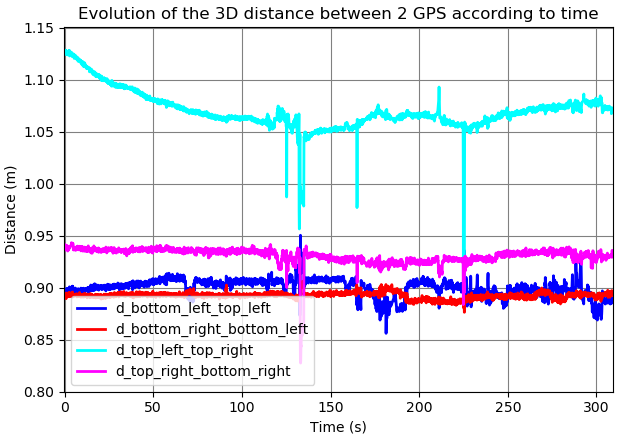}
\caption{Distances between 4 RTK-GPS: case without motion and with disruptions.} \label{disruptions}
\end{figure}

\begin{figure}[htbp]
\centering
\includegraphics[scale=0.4]{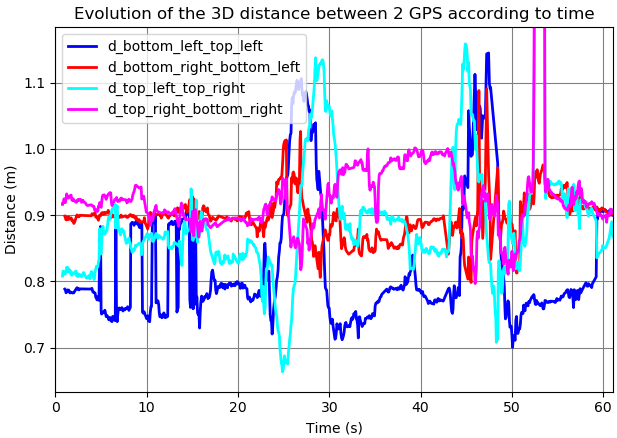}
\caption{Distances between 4 RTK-GPS: a clockwise rotation (one 360-degree turn) is applied to the board, in one direction, then the other. At 54 seconds, a peak is obtained for \texttt{d\_top\_left\_top\_right} and \texttt{d\_top\_right\_bottom\_right} at 1.4 and 1.5 meters respectively.} \label{rotate}
\end{figure}

\begin{figure}[htbp]
\centering
\includegraphics[scale=0.4]{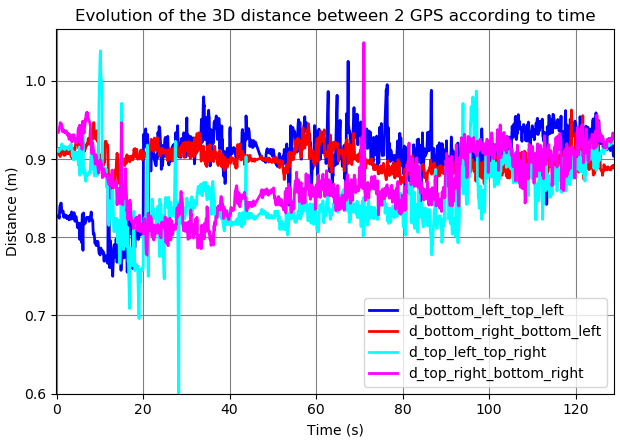}
\caption{Distances between 4 RTK-GPS: translational movement over 130 meters of the board, to form the outline of a square.}
\label{combinaison carre}
\end{figure}

We perform interactions and collaborations between agents, such as following a human on an outdoor path, or detecting a human and estimating its location in the environment. For this, we need centimeter-order accuracy. We propose to carry out an experiment to test the accuracy of an agent's location in space. To do this four RTK GPS rovers are fixed on a board to form a square of 90 cm side (Fig. \ref{rviz_carre}.b). In addition, a base is placed in the middle of a grass field, and the board will be moved in different ways. With our system, we retrieve the 3D position data of the four GNSS antennas using ROS 2 (see Fig. \ref{rosgraph_carre}), particularly using the ROS bag tool to record the data and replay them later. The relative 3D distances between two GPS on one side of the square are calculated and plotted. We can thus visualize the four GPS in 3D in RViz 2 with the local map of the world, then plot the curves obtained in the RQt graph presented in Fig. \ref{plot sur place}, \ref{disruptions}, \ref{rotate}, and \ref{combinaison carre}. We use Emlid Reach M+ with a single-band receiver (L1 at 1 575.42 MHz) and get a frequency of 14 Hz with the satellite constellations.

First of all, the experimental area is a grassy area in the city center, so the buildings interfere with the reception of a GNSS signal. In addition, the weather conditions also influence the reception of a GPS signal. Despite the difficult environmental conditions, it was possible to achieve good results. For all the following results, the time on the abscissa of the graphs is for indication.

We carried out four experiments, illustrated by the four figures obtained. Thus the first test consists in putting the board with the four GPS on the ground and observing the evolution of the 3D distance between two GPS on one side of the square in Fig. \ref{plot sur place}. The four curves represent the four sides of the square as illustrated in Fig. \ref{rviz_carre}.b. 
We observe that we are close to the expected 90 cm and that with time for the four curves, the distance converges to the target value thanks to the accumulation of positions, as can be seen in the dark blue and purple curves. Several experiments have shown that at least one or two curves (in this case, the light blue and red curves) always have an error between 0 and 20 cm. We are determining the cause of this problem. Nevertheless, the curves are stable over time, and the error does not increase or decrease: the location is \textbf{stable} over time, so the error correction is easy.

The second experiment in Fig. \ref{disruptions} also involves leaving the board on the floor, and this time applying a few disturbances. As in the previous test, the values converge towards 90 cm during the first 120 seconds: with no movement or disturbance, very high precision is achieved. Next, we twist the board by lifting the top right corner (twisting it) at 140, 160, and 230 seconds. Peaks are observed involving the light blue and magenta curves in relation to the GPS in the top right corner. Disturbances last only a few seconds, and the error produced is about ten centimeters. Finally, a second type of disturbance is created by rapidly passing our hand over the top left GNSS antennas between 170 to 220 seconds, and 235 to 300 seconds. The position seems to be \textbf{robust} to slight perturbations and we always keep a relative accuracy below 20 centimeters in all cases.

A first type of displacement in Fig. \ref{rotate} is to take the board which is of course parallel to the ground, and apply a rotation in one direction, followed a few seconds later by a rotation in the other direction. The board is carried by two humans on each side. Different phases can be observed, the first 20 seconds when the board is lifted from the ground, then a clockwise rotation is applied (one 360-degree turn), a break without movement is taken from 33 to 43 seconds, and finally, we turn in the other way (one 360-degree turn again) from 43 to 50 seconds. Then, when the board is back on the ground, we note that an error is present (a peak), corresponding to a bad manipulation where a person's hand was passed over the top right GNSS receiver for 2-3s. We conclude with several points. Firstly, throughout the manipulation, the four distances of the square are more or less correct to within 20 cm, despite having lifted the board and turned it rapidly in both directions, all in the presence of two humans (obstructing reception) and in just 60 seconds. Furthermore, even if one of the receivers is obstructed for a few seconds, there is no loss of signal reception, and the system is accurate to within a meter. So the system \textbf{quickly responds} to any disturbance.

Finally, we carried out a normal movement which is as close as possible to the reality of an agent's motions: we walk while holding the board in a straight line for about 30 meters (0 to 45 s) to perform a first translation, a stop is made for 10 seconds, then we finally form the outline of a square on the ground with three translations over a hundred meters and obtain the results Fig. \ref{combinaison carre}. Despite an experiment lasting over 2 minutes, a few small peaks are observed, and we obtained a \textbf{relative accuracy below 20 cm}. Our moves are visible with RViz 2 in Fig. \ref{rviz_carre}.a where the tracks left during the motion can be seen. The square on the path displayed in 3D is \textbf{highly representative of the actual movement achieved} by the two humans walking and holding the board. The four trails left by the four GPS receivers show the sides of the square and the bends. We note that the people carrying the board have returned to the starting point and passed it by about one meter, which is well obtained in our 3D display.

Thanks to these four experiments and other additional tests (not shown in this paper due to space limitation), we've learned several lessons. To improve GPS signal reception, the horizon should be as clear as possible, with few obstacles nearby. Good weather conditions are also important. Environmental conditions can complicate the reception of GNSS signals and decrease positioning accuracy. Also, the system is highly accurate when no movements or disturbances are applied and converges to the expected value. In addition, we observed that the tested RTK-GPS is robust against disturbances and reacts quickly to external perturbations. The designed system should tolerate short term transitive disturbances. The 3D spatial representation of movements has been seen very faithful to reality. Finally, we conclude from the curves and the previous tests, that during continuous movement including rotations and translations, the relative accuracy for positioning agents using RTK technology is accurate within +/-20 cm, which is acceptable for most of the real-world HMAS.

\section{Conclusion}

In this paper, we designed and built a heterogeneous multi-agent system (HMAS) using the ROS 2 middleware. We realized the system architecture and in particular defined how to represent an agent in our HMAS and integrate it. In order to conduct interactions and collaborations, agents are precisely located using RTK GPS and thus expressed in the same referential. We have shown, through both the implementation of a testbed and the use of RTK GPS, that our proposed ROS 2 based HMAS software architecture allows us to seamlessly integrate heterogeneous robots and precisely locate agents (robots and human operators) in the real environment. So the developed ROS 2 based HMAS framework contributes to decreasing the complexity when integrating and making collaborate heterogeneous robots.

\bibliographystyle{IEEEtran}

\bibliography{mybibliography}

\end{document}